\documentclass[twocolumn,prb, showpacs]{revtex4}
\usepackage{graphicx}
\usepackage{dcolumn}
\usepackage{bm}

\begin{document}


\title{Staging model of the ordered stacking of vacancy layers and phase separation in the layered Na$_x$CoO$_2$ ($x \gtrsim$ 0.71) single
   crystals}

\author{G. J. Shu$^{1}$}
\author{F. -T. Huang$^{1,2,3}$}
\author{M. -W. Chu$^{1}$}
\author{J. -Y. Lin$^4$}
\author{Patrick A. Lee$^{5}$}
\author{F. C. Chou$^{1,6}$}
\email{fcchou@ntu.edu.tw}

\affiliation{
$^1$Center for Condensed Matter Sciences, National Taiwan University, Taipei 10617, Taiwan}
\affiliation{
$^2$Taiwan International Graduate Program, Academia Sinica,Taipei 10115,Taiwan}
\affiliation{
$^3$Department of Chemistry,National Taiwan University,Taipei 10617,Taiwan}
\affiliation{
$^4$Department of Physics, National Jiao-Tong University, HsinChu 30076, Taiwan}
\affiliation{
$^5$Department of Physics, Massachusetts Institute of Technology, Cambridge, MA 02139, USA}
\affiliation{
$^6$National Synchrotron Radiation Research Center, HsinChu 30076, Taiwan}
\date{\today}

\begin{abstract}
Phase diagram of Na$_x$CoO$_2$ (x $\gtrsim$ 0.71) has been reinvestigated using electrochemically fine tuned single crystals.  Both phase separation and staging phenomena as a result of sodium multi-vacancy cluster ordering have been found.  Phase separation phenomenon is observed in the narrow ranges of  0.76 $\lesssim$ x $\lesssim$ 0.82 and 0.83 $\lesssim$ x $\lesssim$ 0.86.  While x = 0.820 shows A-type antiferromagnetic (A-AF) ordering below 22K, x = 0.833 is confirmed to have a magnetic ground state of A-AF ordering below $\sim$8K and is only reachable through slow cooling.  In addition, x = 0.859 is found to be responsible for the highest A-AF transition temperature at about 29K.  Staging model based on ordered stacking of multi-vacancy layers is proposed to explain the hysteretic behavior and A-AF correlation length for x $\sim$ 0.82-0.86.
\end{abstract}



\maketitle

\section{\label{sec:level1}Introduction\protect\\ }

Layered material Na$_x$CoO$_2$ has a rich electronic and magnetic phase diagram as a function of x, from A-type antiferromagnetic ordering for x $\gtrsim$ 0.75, to metal-to-insulator transition for x $\sim$ 1/2, and even superconductivity for x $\sim$ 1/3 after hydration.\cite{Foo2004}  Although A-type AF magnetic ordering transition below 22K has been reported in all samples of nominal x from 0.75 to 0.85, the difference among these concentrations has usually been ignored, either due to poorly controlled Na level from melt growth or roughly estimated Na content.\cite{Wooldridge2005, Mendels2005, Bayrakci2005} The high Na vapor loss during high temperature melt growth is well known and the diffusive nature of Na ions at room temperature makes the control of Na content even more difficult,\cite{Shu2007} which can often lead to an inhomogeneous mixture of phases for x $\gtrsim$ 0.7.  Only until recently, detailed Na ion ordering has been revealed through neutron and synchrotron X-ray diffraction studies on single crystal samples.\cite{Roger2007, Chou2008}  The newly found evidence of superstructure formed by multi-vacancy clusters in x $\sim$ 0.71 and 0.84 introduces an idea of doped holes partial localization, which is able to resolve many intriguing physical phenomena found in this layered system, including the Curie-Weiss behavior of a metallic system, the enhanced thermoelectric power, the novel spin liquid state, the reconstructed Fermi surface, and the origin of A-type AF ordering.\cite{Balicas2008, Lee2006, Chou2008} Recent studies of x $\sim$ 0.80 and 0.85 by Schulze $\textit{et al.}$ conclude that Na ordering is highly dependent on the cooling rate, where an additional magnetic ordering below 8K appears only after the sample is slowly cooled through the 300-200K range.\cite{Schulze2008}  However, the real impact of the successive Na rearrangement processes remains to be clarified and the phase diagram must be revisited.




\begin{figure}
\begin{center}
\includegraphics[width=3.5in]{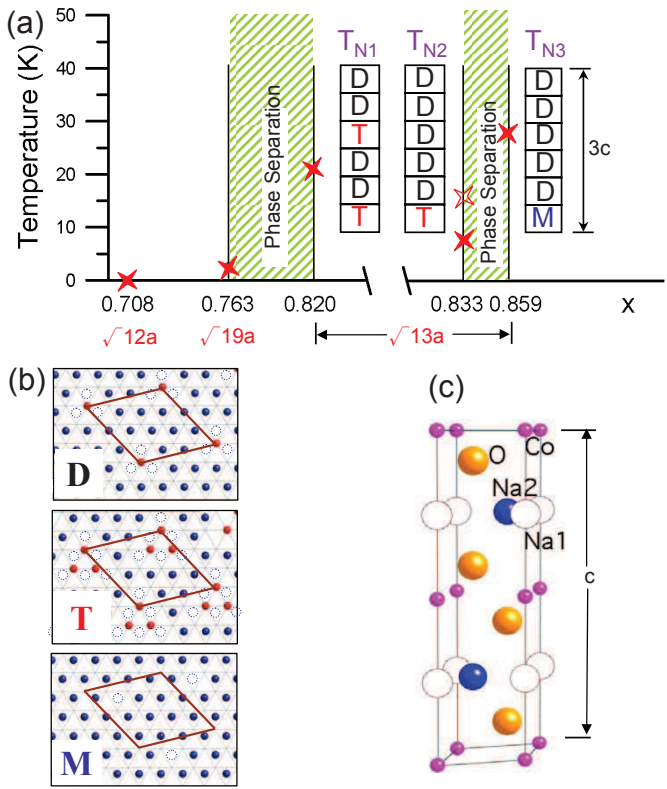}
\end{center}
\caption{\label{fig:fig1}(color online) (a) A revised phase diagram of Na$_x$CoO$_2$ in the range of 0.71-0.86 and the proposed staging models for x=0.820(T$_{N1}$=22K), 0.833(T$_{N2}$=8K) and 0.859(T$_{N3}$=29K). Magnetic ordering temperatures are marked by red cross.  (b) Na layers with tri-vacancy (T), di-vacancy (D) and mono-vacancy (M), where Na2 ions (blue) move from the original Na2 site (empty circle) to the Na1 site (red).  (c) Crystal structure for $\gamma$-Na$_x$CoO$_2$ of P6$_3$/mmc symmetry with vacant Na1 sites shown in empty circles.} 
\end{figure}

Herein, using results from additionally improved electrochemical techniques, specific heat and high resolution single crystal synchrotron X-ray Laue diffraction, we report detailed magnetic and structural phase diagram in the region of 0.71 $\lesssim$ x $\lesssim$ 0.86.  Structural phase separation phenomenon is found in the two regions of 0.76-0.82 and 0.83-0.86 at room temperature as shown in Fig.~\ref{fig:fig1}.  
 While simple hexagonal superstructure of $\sqrt{13}$a is maintained in all samples with x in the range of 0.82-0.86, the magnetic ground state turns out to be distinctively different.  In fact, there are three distinct A-AF transition temperatures of T$_{N1}$=22K, T$_{N2}$=8K and T$_{N3}$=29K found in this range, corresponding to a proposed specific multi-vacancy layer stack ordering of well-defined stoichiometry of x = 0.820, 0.833, and 0.859 respectively, plus x=0.763 that shows a spin glass like behavior below $\sim$ 3K and with a significantly larger superlattice of $\sqrt{19}$a.  In particular, we find x=0.820 to be the most stable phase and cooling rate independent, while x=0.833 shows a strong cooling rate dependent nature with transitions found near 8K (through slow cooling) and 16K (through fast cooling).  These distinctively different A-AF phases found in 0.820, 0.833 and 0.859 can be reasonably constructed by adding different levels of additional Na vacancy to the ideal di-vcancy formed x=11/13=0.846 superstructure of $\sqrt{13}$a$\times$$\sqrt{13}$a$\times$3c.\cite{Chou2008, Huang2009}  Applying a layered staging model similar to that used in graphite intercalated compounds (GIC),\cite{Dresselhaus2002} for example, T$_{N1}$ phase (x=0.820) can be described as a stage-2 compound, i.e., where correct stoichiometry is obtained by introducing two more Na vacancies into the original ideal $\sqrt{13}$a$\times$$\sqrt{13}$a$\times$3c super unit cell, and these defects create tri-vacancy layers that are sandwiched between every two di-vacancy layers. On the other hand, T$_{N2}$ phase (x=0.833) corresponds to stage-5, i.e., tri-vacancy layers are sandwiched between every five di-vacancy layers.  
Most interestingly, the staging model suggests that the trivacancy layers serve as nucleation centers for the interlayer AF ordering. This picture naturally explains why x=0.820 has a higher T$_N$ than that of x=0.833 because of its shortest inter-trivacancy layer distance.  The observed phase separation phenomenon is a natural consequence of the competing multi-vacancy cluster size, superlattice size, and interlayer magnetic correlation.


\section{\label{sec:level1}Experimental\protect\\}
High quality single crystals of well controlled Na content were prepared using electrochemical de-intercalation technique starting from high Na content crystal of x $\sim$ 0.84 grown with floating-zone method, the details have been documented previously.\cite{Shu2007, Shu2008, Chou2008}  The exact Na content has been cross checked with c-axis vs. x linear relationship constructed from combined high angle X-ray diffraction (008) peak position, Inductively Coupled Plasma (ICP) and Electron Probe Microanalysis (EPMA) techniques.\cite{Foo2004, Shu2007, Chou2008}   In particular, current study uses c(x) linear function that is further calibrated by the phase separated boundaries and EPMA is averaged out from freshly cleaved crystal surface for more than 100 points.  A complete list of samples studied is summarized in Table~\ref{tab:table-x}.  
Due to the active diffusive nature of Na ions at room temperature and the minute differences of Na content, all measurements were done on freshly prepared crystals within days.  Otherwise crystal samples must be stored within L-N2 dewar below 200K in order to suppress Na loss from the surface.  Synchrotron Laue diffraction for Na superstructure investigation was performed with synchrotron source in Taiwan NSRRC, and magnetic property characterization was done using Quantum Design SQUID MPMS-XL.

\section{\label{sec:level1}Results and Discussions\protect\\ }

\begin{figure}
\begin{center}
\includegraphics[width=3.5in]{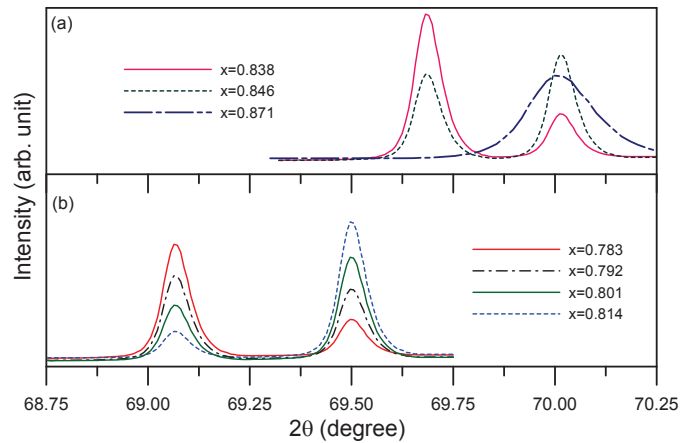}
\end{center}
\caption{\label{fig:fig2}(color online) (a) X-ray diffraction results of (008) peak for samples with x in the ranges of (a) 0.83-0.87 and (b) 0.76-0.82 at room temperature. The linear x-dependence of relative change of (008) diffraction integrated intensities suggests the phase separation phenomenon.  The broadened diffraction peak for x $>$ 0.86 as shown in (a) indicates poor ordering. }
\end{figure}

\begin{table}
\caption{\label{tab:table-x} Summary of studied Na$_x$CoO$_2$ crystal samples}
\begin{tabular}{ccccccc}
 \hline
sample \# & 1-5 & 6-10 & 11-15 &  16-20 & 21-25 &  \\
 \hline
x (EPMA) & 0.768(4) & 0.801(5) & 0.820(3) & 0.832(2) & 0.842(5) \\
x (c-axis) & 0.771(2) & PS & 0.820(2) & 0.832(2) &  PS \\
x (model) & 0.763 & PS & 0.820 &  0.833 & PS & 0.859 \\
T$_N$(K)* & 3** & 22 & 22 &  8 & 8/29  \\
\hline
\end{tabular}
*slow cooling ~ **spin glass like behavior 
\end{table}

While zooming in the region of x $>$ 0.75 using electrochemical technique, we found several concentrations to be particularly stable and of two-phase character.  As indicated in Fig.~\ref{fig:fig2}, the evidence of phase separation is demonstrated by the x-dependent evolution of (008) diffraction peak integrated intensities, where the growth of one end phase is at the expense of the other at the miscibility gap boundaries without continuous intermediate phase in between.   As shown in Fig.~\ref{fig:fig1} and Fig.~\ref{fig:fig2}, phase separation phenomenon is found to occur in two regions of x $\sim$ 0.76-0.82 and 0.83-0.86, where coexistence of two end phases grow at the expense of each other as indicated by the (008) X-ray diffraction peaks, while solid solution connects phases between miscibility gaps.    Clearly the growth of 0.82 phase is at the expense of 0.76 phase for increasing x.  Such coexisting pattern of two phases can only be observed in the narrow regions of 0.76-0.82 and 0.83-0.86, although O3-type secondary phase is commonly found for x $\gtrsim$ 0.85 from melt growth.\cite{Shu2008, Lee2006}  Pure x$\sim$0.86 phase cannot be prepared using electrochemical technique, partly due to the fact that it becomes harder for Na to intercalate into the host structure of compressed c-axis.\cite{Shu2007, Chou2008}  Enforcing more Na into the matrix electrochemically destroys its ordering as indicated by the broadened diffraction peaks that correspond to x$\gtrsim$0.86 as shown in Fig.~\ref{fig:fig2}(a).

\begin{figure}
\begin{center}
\includegraphics[width=3.5in]{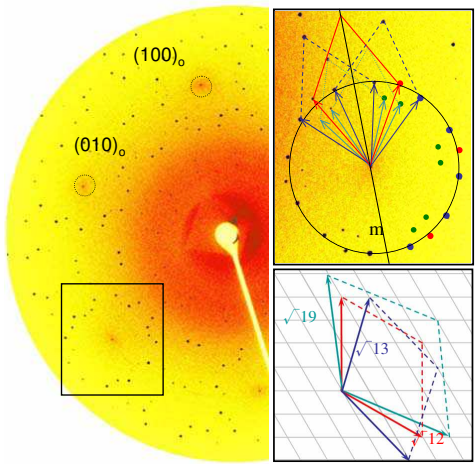}
\end{center}
\caption{\label{fig:fig3}(color online) 
Synchrotron X-ray Laue pattern for x=0.801(5) at room temperature, where three twin sets of hexagonal units are shown in the reciprocal space (upper inset), which correspond to hexagonal $\sqrt{13}$a (blue), $\sqrt{12}$a (red) and $\sqrt{19}$a (green) superlattices in the real space (lower inset). }
\end{figure}

Fig.~\ref{fig:fig3} shows synchrotron X-ray Laue diffraction results for x = 0.801(5), where x sits in the middle of the miscibility gap of 0.76-0.82.  Although in-house x-ray results for x within 0.76-0.82 range all display two-phase feature of the two end compounds only, the synchrotron X-ray Laue shows there are actually three coexist phases at room temperature.   As shown in Fig.~\ref{fig:fig3}, we can clearly identify three twin sets of hexagonal superlattices that correspond to  $\sqrt{19}$a, $\sqrt{13}$a and $\sqrt{12}$a in the real space, the latter two can be compared with the published single phase Laue patterns of x = 0.71 ($\sqrt{12}$a) and 0.84 ($\sqrt{13}$a).\cite{Chou2008, Huang2009}  The newly found $\sqrt{19}$a superlattice must be due to Na ordering for x $\sim$ 0.76 and has been verified by single phase sample, where more Na vacancies are introduced and a larger superlattice becomes necessary to accommodate  larger multi-vacancy clusters.  In fact it requires 4.5 vacancies per each superlattice of size $\sqrt{19}$a to account for stoichiometry of x $\sim$ 0.76, i.e. x = 1 - $\frac{4.5}{19}$ = 0.763 that is composed of quadri- and penta-vacancy clusters in adjacent layers for $\gamma$-Na$_x$CoO$_2$.\cite{Chou2008, Huang2009}  A detailed analysis for x $\sim$ 0.76 sample will be presented elsewhere.  By treating this layered material to be a pseudo-binary system composed of Na and CoO$_2$, the triple coexisting superlattices at constant temperature and pressure does not violate the phase rule, in fact it has reached the allowed maximum number of three.  Considering dominant domains are from x$\sim$0.82 ($\sqrt{13}$a with di-vacancy clusters) and x$\sim$0.76 ($\sqrt{19}$a with quadri/penta-vacancy clusters), it's reasonable to have a buffered zone at the domain boundary which is built with  tri/quadri-vacancy clusters of $\sqrt{12}$a superlattice. 
\begin{figure}
\begin{center}
\includegraphics[width=3.5in]{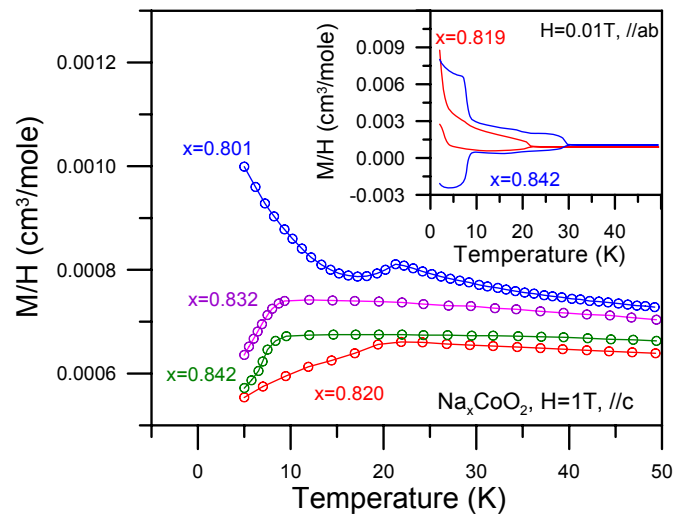}
\end{center}
\caption{\label{fig:fig4}(color online) Magnetic susceptibility measurement results for x $>$ 0.80 under applied field of 1 Tesla and 0.01 Tesla (inset) along the c- and ab-directions respectively.  All measurements were performed after being zero-field-cooled from room temperature through 200K with cooling rate of 2 K/min. A-type AF ordering is indicated by the cusp of c-axis susceptibility $\chi_c$ and the onsets of low field hysteresis shown in the inset.}
\end{figure}

Most of the magnetic susceptibility measurements for Na$_x$CoO$_2$ with x $\gtrsim$ 0.75 before show A-type AF signature near 22K, i.e., the cusp of $\chi_{c}$ under high field.  Although crude magnetic phase mappings in this range before suggest that T$_N$ varies between 22-27K,\cite{Mendels2005, Sugiyama2004} Schulze $\textit{et al.}$ recently found an additional 8K phase for x $\sim$ 0.80 and 0.85, which can be obtained only after a slow cooling process.\cite{Schulze2008}  With carefully tuned single crystals in the narrow range of 0.82-0.86, we are able to re-visit the magnetic phase diagram and untangle the mystery of T$_N$ variation.  Magnetic susceptibility measurement results are shown in Fig.~\ref{fig:fig4}, all measurements were done after a slow cooling rate of 2 K/min.   We find that T$_N$ does not change with x monotonously and continuously, instead, the four phases at the two miscibility gap boundaries shown in Figs.~\ref{fig:fig1} and \ref{fig:fig4} are responsible for the different characteristic T$_N$'s, where x $\sim$ 0.76 shows Curie-Weiss behavior down to 5K (not shown).  The onsets of A-AF transitions are indicated by the cusp of $\chi_c$ under high filed, which occur at T$_{N1}$=22K, T$_{N2}$=8K and T$_{N3}$=29K for x = 0.820, 0.833 and 0.859 respectively, while 0.801(5) and 0.842(5) data reflect their mixed phase nature, i.e., superposition of contributions from the end members of 0.76-0.82 and 0.83-0.86 miscibility gaps respectively.   There is ZFC/FC irreversibility found below T$_N$ for both $\chi_c$ and $\chi_{ab}$ at low field, although stronger FM saturation moments are seen along the ab-direction.  Such A-AF ordering has been verified by the neutron scattering for Na$_{0.82}$CoO$_2$,\cite{Bayrakci2005} where strong field dependence of magnetization below T$_N$ $\sim$ 22K has been confirmed to be metamagnetic.\cite{Luo2004}  Current low field measurement is in agreement with that reported for x $\sim$ 0.85,\cite{Schulze2008} although our data indicate that the 8K phase is coming from the phase of x closer to 0.833, while a more stable phase of 22K is from x closer to 0.820.

\begin{figure}
\begin{center}
\includegraphics[width=3.5in]{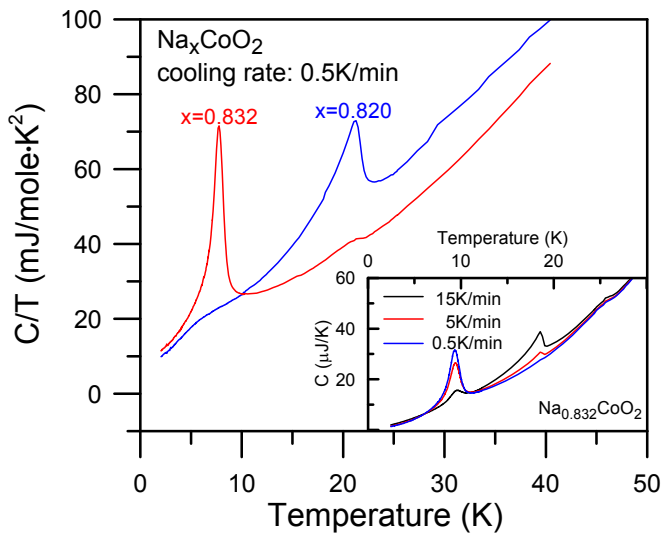}
\end{center}
\caption{\label{fig:fig5}(color online) Specific heat measurement results for x = 0.820(3) and 0.832(2) obtained from slowest cooling rate of 0.5 K/min.  The inset shows x=0.832(2) has a strong cooling rate dependence, where 8K phase grows at the expense of 16K phase as cooling rate reduces from 15 to 0.5 K/min.}
\end{figure}

We find that the different onsets of A-AF transition between x = 0.820(3) and 0.832(2) are clearly demonstrated by the cooling rate dependence of T$_N$ as revealed by the specific heat data shown in Fig.~\ref{fig:fig5}, where 22K transition for x=0.820(3) is independent of cooling rate, while fast cooling rate moves T$_N$ discretely from 8K to 16K for x=0.832(2).  16K phase occurs in x=0.832(2) when a fast cooling is applied, while it decreases at the expense of 8K phase generation under decreasing cooling rate, although the existence of minor 22K is difficult to avoid completely for samples near the phase separated boundary.  The ratio of the minor 22 K phase to the major 8 K phase in the present x=0.833 sample can be estimated to be 3.6$\%$ from the entropy change of the small anomaly at 22 K for Na$_{0.833}$CoO$_2$.  Sodium ion diffusion is active at room temperature for high x samples,\cite{Shu2007} but it freezes below $\sim$200K as indicated by the sharp increase of 1/T$_1$ for $^{23}$Na due to Na motion.\cite{Yoshimura2007} Sample of x=0.832(2) must be cooled through the temperature range of 300-200K with a rate slower than 10 K/min in order to reach the magnetic ground state that corresponds to 8K magnetic ordering.


The entropy associated with the 22 K transition in Na$_{0.820}$CoO$_2$ and the 8 K transition in Na$_{0.833}$CoO$_2$ are estimated to be $\triangle$S$\sim$170 mJ/mol K and $\triangle$S$\sim$200 mJ/mol K, respectively. Taking the surface value of estimated $\triangle$S and the transition width character, these results indicate that the magnetic moments order better in Na$_{0.833}$CoO$_2$ than in Na$_{0.820}$CoO$_2$.   The relatively poor ordering of stage-2 for x=0.820 (T$_{N1}$=22K) than that of stage-5 x=0.833 (T$_{N2}$=8K) is interestingly in agreement with the alternating T-Q stacking requirement as described by the x=0.71 superstructure model before,\cite{Huang2009, Chou2008} i.e., tri-vacancy is not favorable in the even-layer within P6$_3$/mmc symmetry and there must exist mixing stages of 1 and 3 for x=0.820.  The $\triangle$S value for x=0.833 is more than 10 times larger than that reported in Ref. [10], which suggests a nearly single 8 K phase in the present sample.  On the other hand, these measured $\triangle$S values are only about 20$\%$ of the entropy estimated from the complete ordering of fully localized spin-1/2 Co$^{4+}$ ions.  This discrepancy might indicate the failure of the simple ionic Co$^{3+}$-Co$^{4+}$ picture.  

\begin{figure}
\begin{center}
\includegraphics[width=3.5in]{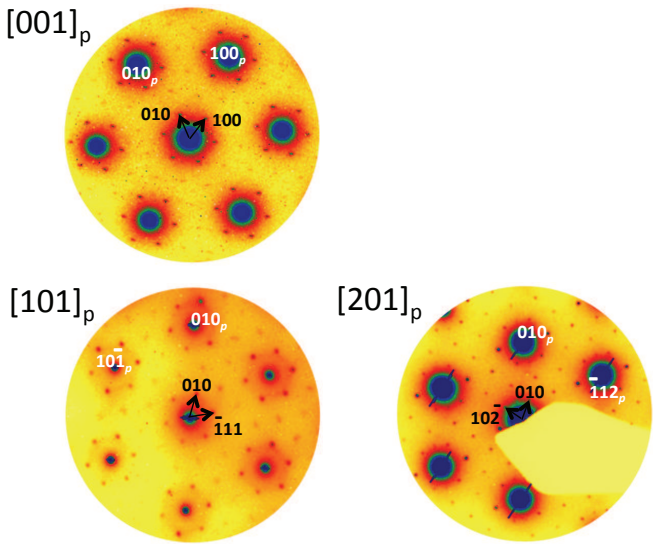}
\end{center}
\caption{\label{fig:fig6}(color online) ) Electron diffraction patterns of single domain Na$_{0.820}$CoO$_2$ for transmitted e-beam along [001]$_p$,  [101]$_p$, and [201]$_p$ projections (p, denoting the primitive lattice). The superlattice spots surrounding the intense primitive cell reflections in the ab-plane projection, [001]$_p$, indicate the superlattice ordering of $\sqrt{13}$a$\times$$\sqrt{13}$a without c-axis information.  In the azimuthal projections of  [101]$_p$ and [201]$_p$, the superlattice reflections suggests 3c ordering, i.e. can only be indexed correctly with  $\sqrt{13}$a$\times$$\sqrt{13}$a$\times$3c superlattice.}
\end{figure}

Since all samples with x $\gtrsim$ 0.82 show identical superlattice size of $\sqrt{13}$a, the subtle difference for x $\sim$ 0.82 and 0.83 must be related to the Na rearrangement generated by the additionally introduced Na defect that causes deviation from the ideal di-vacancy constructed 0.846=11/13 of $\sqrt{13}$a$\times$$\sqrt{13}$a$\times$3c superstructure.\cite{Chou2008, Huang2009}  But what kind of mechanism is responsible for these discrete T$_N$'s of $\bigtriangleup$x only 1-3$\%$ apart?  The secret lies in the stack ordering of 2D hexagonal superlattices.  From our previous studies on the structure of 0.71 and 0.84,\cite{Chou2008, Huang2009} the ideal superlattice has a 3c periodicity.  The 3c periodicity for x=0.820 is once again confirmed by electron diffraction patterns on single domain crystals as shown in Fig.~\ref{fig:fig6}.  Although [001]$_p$ diffraction pattern cannot tell the periodicity along c-direction, perfect indexing for diffraction patterns with transmitted beam along primitive [101]$_p$ and [201]$_p$ can only be achieved with the help of 3c periodicity assignment.  When one and two more Na defects per 3c unit (i.e., six layers of Na) are introduced into the perfectly ordered original 0.846 = 11/13 superstructure,\cite{Chou2008} two additional stoichiometries of 0.833 = 0.846$-\frac{1}{6}\times\frac{1}{13}$ and 0.820 = 0.846$-\frac{2}{6}\times\frac{1}{13}$ can be introduced, as verified by our X-ray and magnetic measurement results shown above.  When magnetic ordering occurs, spins from itinerant electron or localized electrons near Co ions are certain to be affected by the rearrangement of Na multi-vacancy clusters in the nearby layers.  In fact the in-plane inter-vacancy cluster distance $\sqrt{13}$a is nearly twice the inter CoO$_2$ distance.  Since every one extra Na vacancy introduced into the ideal 3c unit would convert the original layer of di-vacancy formed 2D superstructure into tri-vacancy, we can thus simplify the stacking problem into stack ordering between the di-vacancy (D) and tri-vacancy (T) layers along the c-direction.  We propose a new staging model  as shown in Fig.~\ref{fig:fig1} to explain such stack ordering, which shows strong resemblance to the staging phenomenon often observed in the 2D intercalated graphite compounds.\cite{Dresselhaus2002}  The x=0.833 phase which has only one Na defect introduced could have a stage-5 construction, while x=0.820 phase of two Na defects per 3c unit must have a stage-2 construction, i.e., there are five and two D-layers sandwiched in between T-layers.

We can now use the staging picture to interpret the variety of magnetic ordering temperatures observed in the range 0.82 $\lesssim$ x $\lesssim$ 0.86. As discussed previously,\cite{Chou2008} the di-vacancy may localize a carrier on the adjacent Co layer, leaving a low density hole gas of density $\frac{1}{13}$, which is unstable to Stoner ferromagnetism. This may be the origin of the ferromagnetic layers which then order antiferromagnetically between layers to form the A type AF ordering.  Here we suggest that the driving force for interlayer coupling may lie in the T-layer. The tri-vacancy has one extra negative charge which lowers the tunneling barrier between the  localized holes on the adjacent layers and enhances the antiferromagnetic spin correlation between them. Thus the T-layer may form the nucleation layers to drive the three dimensional AF order. This picture explains why T$_{N1}$ is 22K for x=0.820 vs. T$_{N2}$ is 8K for x=0.833 where the spacing between T-layers is much larger. Upon rapid cooling, some stage 4 and stage 6 states may form. The stage 4 meta-stable phase may be responsible for the intermediate T$_N$ of 16K. The hysteretic behavior observed below T$_N$ (see Fig.~\ref{fig:fig4}) could also be explained by either the in-plane FM domain effect or by uncanceled A-type AF moments along the c-axis as a result of mixed staging. The phase separation observed near 0.83-0.86 can also be explained using the same stage model.  Since 0.859=0.846+$\frac{1}{6}\times\frac{1}{13}$, i.e., one more Na ion (not vacancy) is introduced into the original ideal x=0.846 phase of $\sqrt{13}$a$\times$$\sqrt{13}$a$\times$3c superstructure, the di-vacancy is converted to a mono-vacancy (M) forming a stage 5 stacking.  The hole density on either side of the M-layer is now reduced by $\frac{1}{2}\times\frac{1}{13}$ and we may expect an even strong tendency toward Stoner ferromagnetism. The higher transition temperature of the Co layers adjacent to the M-layer may explain the higher T$_{N3}$ = 29K for x=0.859.

\section{\label{sec:level1}Conclusions\protect\\ }

In conclusion, we have revised Na$_x$CoO$_2$ phase diagram in the range of 0.71-0.86 using electrochemically fine tuned single crystal samples.  The puzzling and inconsistent measurement results in this range before have been clarified and interpreted as a result of phase separation and staging phenomena.  A direct link between the high temperature Na ion (vacancy) ordering  and the low temperature magnetic properties have been established.  The highly correlated ion, magnetic and charge orderings in layered Na$_x$CoO$_2$ can provide invaluable information to the study of strongly correlated electron low dimensional system which has itinerant electrons on a triangular lattice.

\section*{Acknowledgment}
FCC acknowledges the support from National Science Council of Taiwan under project number NSC 97-3114-M-002.  PAL acknowledges support by the DOE grant number DE-FG02-03ER46076.


\end{document}